\documentstyle[12pt]{article}
\input psfig
\begin{document}
\thispagestyle{empty}
\null
\begin{flushright}
UFIFT-HEP-96-14
\end{flushright}
\vspace{.4in}

\begin{centering}

{\LARGE\bf What is the spectrum of cold dark\\ 
\vspace{.1in}
matter particles on Earth?$^{\dagger}$} \\
\vspace{.75in}

{\large Pierre Sikivie}\\
\vspace{.2in}

{\it Department of Physics, University of Florida}\\
{\it Gainesville, FL 32611, USA.}\\

\date{ }
\vspace{.75in}

\end{centering}
\vspace{.1in}

\begin{abstract}
It is argued that the spectrum of cold dark matter particles
on Earth has peaks in velocity space associated with particles
falling onto the Galaxy for the first time and with particles 
which have fallen in and out of the Galaxy only a small number
of times in the past.  Estimates are given for the sizes and
velocity magnitudes of the first few peaks.  The estimates are
based upon the secondary infall model of halo formation which 
has been generalized to include the effect of angular momentum.
\end{abstract}

\vspace{2.5in}
\footnoterule

\noindent {\footnotesize $^{\dagger}$Invited talk at the DM96 Workshop
on Sources and Detection of Dark Matter in the Universe,
Santa Monica, CA, Feb. 14--16, 1996}

\newpage

\section*{\large\bf 1.\ \ INTRODUCTION AND OVERVIEW}
         This talk is based entirely upon work done in collaboration
with J. Ipser\cite{ipser1} and with I. Tkachev and Y. Wang\cite{tkac}.
	
	Experiments are under way which attempt to identify
the nature of dark matter by direct detection on Earth\cite{pfs}.  
The candidates which are being searched for in this way are axions
with mass in the $\mu$eV range and WIMPs (weakly interacting 
massive particles) with mass in the GeV range.  Axions and 
WIMPs are the leading cold dark matter (CDM) candidates.  Other
forms of dark matter are baryons and neutrinos\cite{dark}.  From 
the point of view of galaxy formation, the defining properties 
of CDM are:
\begin{enumerate}
\item that CDM particles, unlike baryons, are guaranteed to
interact with their surroundings only through gravity, and
\item that CDM particles, unlike neutrinos, have negligibly
small primordial velocity dispersion.
\end{enumerate}
Studies of large scale structure support the view that the 
dominant fraction of dark matter is CDM.  Moreover, if some
fraction of dark matter is CDM, it necessarily contributes 
to galactic halos by falling into the gravitational wells of 
galaxies and hence is susceptible to direct detection on Earth.

	Motivated by the prospect that a direct search experiment
may some day measure the spectrum of CDM particles on Earth,
one may ask what can be learned from that spectrum about our galaxy 
and the universe.  In particular, if a signal is found in the 
cavity detector of galactic halo axions\cite{axdet}, it will be 
possible to measure the CDM spectrum with great precision and 
resolution.  In addition, there is the possibility that 
beforehand knowledge of some spectral properties helps 
in the discovery of a signal.
 
	In many discussions of dark matter detection on 
Earth\cite{pfs}, it has been assumed that the dark matter 
particles in our galaxy have an isothermal velocity 
distribution.  A strong argument in support of this assumption 
is the fact that a self-gravitating isothermal sphere always has
a density profile $\rho(r)$ which falls off at large $r$ as 
$1/r^2$ and hence a flat rotation curve.  However, it easy to 
convince oneself that the velocity distribution of dark matter 
particles necessarily has a component which is not isothermal.

	Indeed, consider the fact that our closest neighbor 
on the galactic scale, the galaxy M31 in Andromeda, at a distance
of order 730 kpc from us, is falling towards our galaxy with a line-
of-sight velocity of order 120 km/sec.  This motion can be understood 
to be due to the mutual gravitational attraction between the two 
galaxies.  We may use it as an indicator of the motion of any matter 
in our neighborhood.  Now, if CDM  exists, it is present everywhere 
because, by Liouville's theorem, the 3-dim. sheet in 6-dim. phase-space 
on which the CDM particles lie can not be ruptured whatever its 
evolution may be. The thickness of that sheet is the tiny primordial 
velocity dispersion of the CDM particles.  The implication is that, 
if CDM exists, there are CDM particles falling onto our galaxy 
continuously and from all directions.  The motion of these particles 
gets randomized inside the galaxy by gravitational scattering off 
giant molecular clouds, globular clusters and other inhomogeneities 
but complete thermalization of their velocity distribution occurs 
only after they have fallen in and out of the galaxy many 
times.

	As a result, there are peaks in the velocity distribution
of CDM particles at any physical point in the galaxy\cite{ipser1}.  
One peak is due to particles falling onto the galaxy for the 
first time, one peak is due to particles falling out of 
the galaxy for the first time, one peak is due to 
particles falling in for the second time, and so on.  Estimates 
have been obtained of the sizes and the velocity 
magnitudes (in a reference frame which is not rotating along with
the disk) of these peaks using the secondary infall model of
galactic halo formation\cite{tkac}.  The existing version of 
this model was generalized to allow the dark matter particles 
to have angular momentum.  Indeed, as will be seen below, 
angular momentum has a large effect upon the peak sizes.

\section*{\large\bf 2.\ \ SELF-SIMILAR SECONDARY INFALL}

In the secondary infall model of galactic halo formation\cite{infall}, 
a halo forms and grows around an initial overdensity because dark 
matter keeps falling onto it.  The dark matter is initally receding 
from the overdensity, as part of the general Hubble expansion, but is 
gravitationally attracted to it.  The dark matter is assumed to be 
non-dissipative and have zero initial velocity dispersion.  The 
gravitational potential of the galaxy is taken to be spherically 
symmetric.  Moreover, in the original formulation of the model, 
the dark matter particles are assumed to have zero angular 
momentum with respect to the center of the overdensity and 
they thus move on radial orbits through it.  As is explained 
below, it is possible and, for the purpose of estimating 
velocity peaks, necessary to rid the model of this last 
assumption. But let's keep the assumption for the moment 
for pedagogical purposes.

\begin{figure}[htb]
\psfig{figure=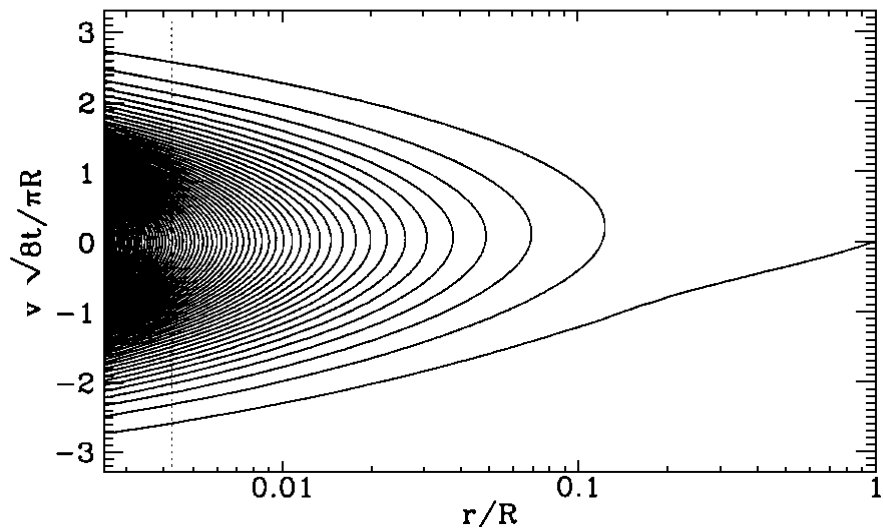,width=4.8in}
\caption{The line is the location of dark matter particles
in phase space at a fixed moment of time in the secondary infall
model with $\epsilon = 0.2$ and $j = 0$.  The dotted line 
corresponds to the Sun's position if $h = 0.7$.}
\end{figure}

An initial overdensity profile $\delta M_i(r)$ is assumed.  
The equations of motion for the radial coordinate $r (\alpha,t)$ 
of each spherical shell ($\alpha$ is a shell label, $t$ is time) 
in the gravitational potential due to all the other shells 
must then be solved for initial conditions given by
the Hubble expansion at some arbitrarily chosen but early time $t_i$:
$\dot r (\alpha, t_i) = H (t_i) \tau (\alpha, t_i)$. \ Much progress in
the analysis of the model came about as a result of the realization that
the evolution of the galactic halo is self-similar\cite{fg} 
provided the initial overdensity has the following scale-free form:
\begin{equation}
{\delta M_i\over M_i} = \left ({M_0\over M_i}\right )^\epsilon
\end{equation}
and provided $\Omega = 1$.  \ $M_i$ and $\delta M_i$ are respectively the
mass and excess mass interior to $r_i$ at the initial time $t_i$.  
$\epsilon$ is a parameter in the range $0 < \epsilon \leq 1$.  \ 
Self-similarity means that the phase-distribution of the 
dark matter particles is time-independent after all distances 
have been rescaled by the overall size $R(t)$ of the galactic 
halo and all masses by the mass $M(t)$ interior to the radius 
$R(t)$.  $R(t)$ is taken to be the ``turn-around'' radius at 
time $t$, i.e. the radius at which particles have zero radial
velocity for the first time in their history (see Fig.~1).  
In particular, the mass-profile of the halo 
$M(r,t) = M(t) {\cal M} \left ({ r\over R(t)}\right )$.  
\ It was shown analytically\cite{fg} that ${\cal M} (\xi) \sim
\xi$ as $\xi \to 0$ if $0 < \epsilon \leq {2\over 3}$.  
\ Thus, in the range $0 < \epsilon \leq {2\over 3}$, the 
model produces flat rotation curves, i.e., it is in accord 
with the main feature of the galactic mass distribution.

\begin{figure}[htb]
\psfig{figure=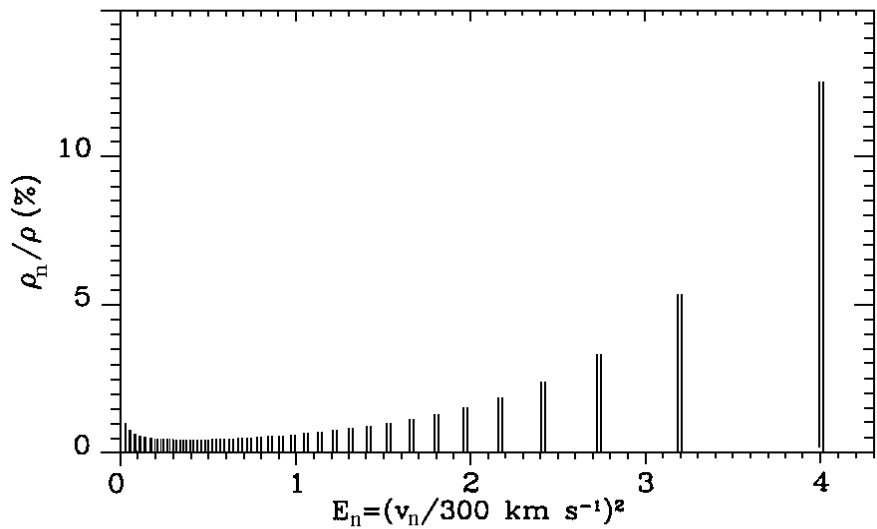,width=4.8in}
\caption{The spectrum of velocity peaks at the Sun's position for the case
$\epsilon = 0.2$, $j = 0$ and $h = 0.7$.}
\end{figure}

\begin{figure}[htb]
\psfig{figure=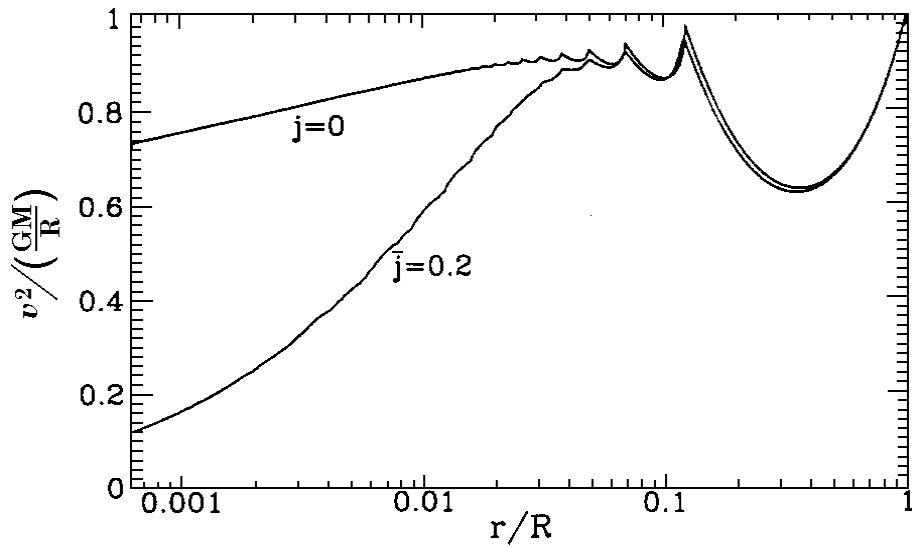,width=4.8in}
\caption{Rotational curves for the case $\epsilon = 0.2$, with and 
without angular momentum.}
\end{figure}

To fit the model to our present galactic halo, one must 
choose appropriate values of $R$ and $M$. \ This was done\cite{tkac} 
by matching the rotation velocity in the model to the observed 
one (220~km/sec) in our galaxy and choosing a value for the age 
of the universe:  $t_0 = {1\over h} (6.52~10^9$~years) 
where $h$ parametrizes the present Hubble rate:  $H_0 = 
h~100$~km~sec$^{-1}$~Mpc$^{-1}$.  \ For typical values of 
$\epsilon$ and $h$, $R$ turns out to be in the 1 to 3~Mpc range.  
Figs.~1 and 2 show the phase-space diagram and the velocity 
peaks on Earth for $\epsilon = 0.2$ and $h = 0.7$.  \ The rows 
labeled $j = 0.0$ in Table~1 give the density fractions and 
kinetic energies of the first five incoming peaks in the
model without angular momentum.  For each incoming peak there is an 
outgoing peak with approximately the same energy and density 
fraction (see Fig.~2).

However, one must question in this context the approximation of
neglecting the angular momentum that the dark matter particles 
are expected to have. The model without angular momentum tends 
to overestimate the size of the peaks due to particles falling 
in and out of the galaxy for the first time.  Indeed, angular 
momentum has the effect of keeping infalling particles away 
from the galactic center and this effect is largest for the 
particles falling onto the galaxy last.  Fortunately, there 
is a generalization of the model which takes angular momentum 
into account while still keeping the model tractable.

\def\dtt{\tt "}

\begin{table*}
\caption{Density fractions $f_n$ and kinetic energies $E_n$ of the
first five incoming peaks for various values of $\epsilon$, 
$\overline j$ and $h$.  Also shown are the current turnaround radius $R$ 
in units of Mpc, the effective core radius $b$ in kpc, and the local 
density $\rho$ in $10^{-25}$~g~cm$^{-3}$.  The $f_n$ are in percent and 
the $E_n$ are in units of $0.5 \times (300 \hbox{km s}^{-2})^2$. }
\vspace{.1in}
{\small
\begin{tabular}{cclclclclclclclclclclclclclclc}
\hline
\hline
$\epsilon$ & $\overline j$ && h &&  R &&  b && $\rho$ && $f_1$ 
($E_1$) && $f_2$ 
($E_2$) && $f_3$ ($E_3$) && $f_4$ ($E_4$) && $f_5$ ($E_5$)\\
\hline 
0.2 & 0.0 && 0.7 && 2.0 && 0.0 && 8.1 && 13 (4.0) && 5.3 (3.2) 
&& 3.3 (2.7) && 2.4  (2.4) && 1.9 (2.2)\\
1.0 & 0.0 && 0.7 && 0.9 && 0.0 && 8.4 && 1.6 (3.4) && 1.1 (3.2) 
&& 0.9 (3.0) && 0.8 
(2.9) && 0.7 (2.8)\\
\hline  
0.15 & 0.2 && 0.7 && 2.4 && 13 && 5.0 && 4.0 (3.1) && 5.4 (2.3) 
&& 5.3 (1.8) &&  4.9 (1.5) && 4.0 (1.3)\\
 0.2 & 0.1 && 0.7 && 2.0 && 4.5 && 7.6 && 7.4 (3.8) && 7.2 (3.0) 
&& 4.9 (2.5) && 3.2 (2.2) && 2.4 (2.0)\\
\dtt & 0.2 && 0.7 && 2.0 && 12 && 5.4 && 3.1 (3.4) && 4.1 (2.6) 
&& 4.3 (2.1) && 4.1 (1.8) && 3.6 (1.6)\\
\dtt & \dtt && 0.5 && 2.8 && 17 && 4.9 && 1.9 (3.5) && 2.5 (2.7) 
&& 2.8 (2.3) && 2.9 (2.0) && 3.0 (1.7)\\
\dtt & \dtt && 0.9 && 1.6 && 9.3 && 6.0 && 4.4 (3.2) && 5.3 (2.5) 
&& 5.1 (2.0) && 4.5 (1.7) && 3.6 (1.5)\\
\dtt & 0.4 && 0.7 && 2.0 && 40 && 2.6 && 0.8 (2.5) && 1.6 (1.8) 
&& 2.1 (1.4) && 2.4 (1.1) && 2.6 (0.9)\\
0.25 & 0.2 && 0.7 && 1.8 && 8.5 && 5.5 && 2.0 (3.5) && 2.9 (2.8) 
&& 3.3 (2.4) && 3.4 (2.1) && 3.1 (1.8)\\
0.4 & 0.2 && 0.7 && 1.5 && 2.2 && 7.7 && 1.1 (4.0) && 1.5 (3.4) 
&& 1.8 (3.0) && 1.9 (2.8) && 2.1 (2.5)\\
\hline
\hline
&\ \\
\end{tabular}
}
\end{table*}

\begin{figure}[htb]
\psfig{figure=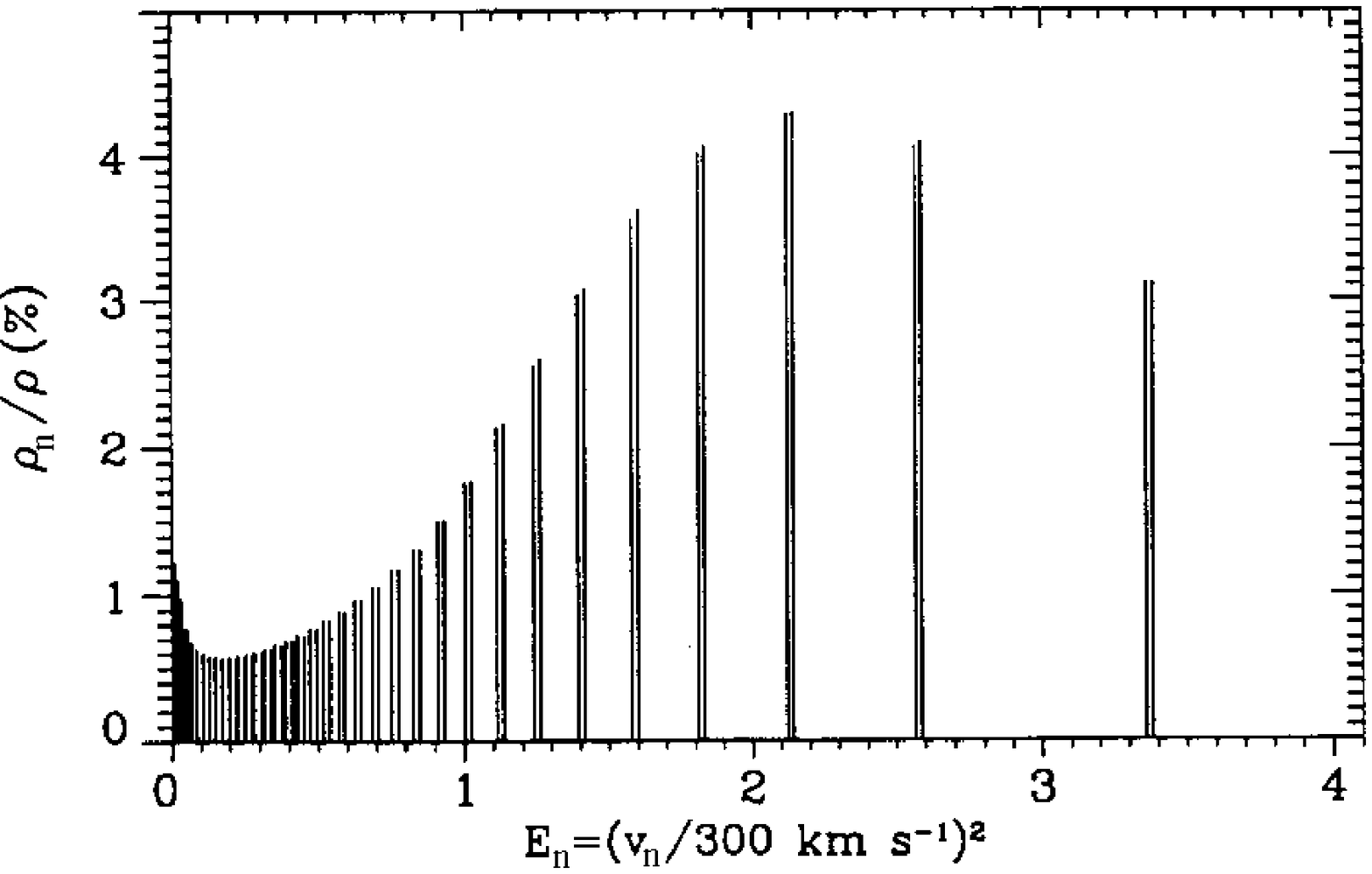,width=4.8in}
\caption{The spectrum of velocity peaks for the case $\epsilon = 0.2$, 
$\overline j = 0.2$ and $h = 0.7$.}
\end{figure}

In this generalization\cite{tkac}, spherical symmetry is maintained
by, in effect, averaging over all possible orientations of 
an actual physical halo.  Each shell $\alpha$ is divided into 
subshells labeled by an index $k$.  \ The particles in a 
given subshell ($\alpha, k)$ all have the same magnitude 
$\ell_k(\alpha)$ of angular momentum.  At any point on a
subshell, the distribution of angular momentum vectors is 
isotropic about the axis from that point to the galactic 
center.  Thus the spherical symmetry of each subshell is 
maintained in time.  Moreover, it was found
that the evolution is self-similar provided:
\begin{equation}
\ell_k (\alpha) = j_k r_* (\alpha)^2/t_* (\alpha)
\end{equation}
where $r_*(\alpha)$ and $t_*(\alpha)$ are the turn-around 
radius and turn-around time of shell $\alpha$ and the  
$j_k$ are a set of dimensionless numbers characterizing the 
galaxy's angular momentum distribution.  In all cases 
presented here, the $j_k$ were taken to be distributed 
according to the density:
\begin{equation}
{dn\over dj} = {2j\over j_0^2} \exp \left( -{j^2 \over j_0^2}\right )\, ,
\end{equation}
for which the average is $\overline j = {\sqrt{\pi} \over 2} j_0$.  \ 
Fig.~3 shows the rotation curves for the cases $\epsilon = 0.2, j = 0$
and $\epsilon = 0.2, \overline j = 0.2$.  \ It shows that 
the effect of angular momentum is to give a core radius to 
the halo, i.e., it makes the halo contribution to the rotation 
velocity go to zero for $ r\to 0$.  \ The `effective core radius' $b$ 
is defined as the radius at which half of the rotation velocity 
squared is due to the halo.

Fig.~4 shows the velocity peaks for the case $\epsilon = 0.2$, 
$\overline j = 0.2$, and $h = 0.7$.  Table~1 gives the values 
of the current turn-around radius $R$, the effective core 
radius $b$, the halo density at our location $\rho$ and 
the density fractions and energies of the five most energetic
incoming peaks for representative values of $\epsilon$, 
$\overline j$ and $h$.  The range of $\epsilon$ values 
chosen is motivated by models of large-scale structure 
formation\cite{ceps} as well as by the fact that the rotation
curve is flat in the range $0 < \epsilon \leq {2\over 3}$.  
\ The table shows that the contribution to the local halo 
density due to particles which are falling in and out of 
the galaxy for the first time or which have passed through 
the galaxy only a small number of times in the past, and 
which are therefore not thermalized, is not small since it 
comprises several percent per velocity peak. 

Finally, let me emphasize that the actual peak sizes on Earth 
and anywhere else in the halo have a probabitity distribution 
which reflects the unknown distribution of angular momenta of 
the infalling dark matter particles.  The peak sizes given in 
Table~1 and Fig.~4 are the {\em average\/} peak sizes for the 
angular momentum distribution of Eq.(3).  More details will be 
given in a forthcoming publication\cite{next}.

\end{document}